\documentstyle[11pt,newpasp,twoside,epsf]{article}
\def\edcomment#1{\iffalse\marginpar{\raggedright\sl#1\/}\else\relax\fi}
\reversemarginpar

\begin{document}
\title{Modification of AGB wind in a binary system}
 \author{Adam Frankowski and Romuald Tylenda}
\affil{Nicolaus Copernicus Astronomical Center, Polish Academy of Sciences,
Rabianska 8, 87--100 Torun, Poland}

\begin{abstract}
We explore one of plausible causes of asymmetry in the wind of an AGB star
-- the presence of a binary companion. We have developed a simple method for
estimating the intrinsic non-sphericity of the outflow in that case. 
Assuming the Roche model, local stellar parameters ($T_{\rm eff}$, $g$) are
calculated, and then a formula for the local mass loss rate is applied.
As a by-product a relation between the ratio of stellar radius to the Roche
lobe radius and the overall wind enhancement is obtained.
\end{abstract}

\section{Introduction}

Shapes of Planetary Nebulae, that deviate from spherical symmetry (in
particular -- axisymmetric ones) are often ascribed to binary interactions
(e.g. Soker 1997).
When an Asymptotic Giant Branch
star loses its mass -- which is to become a PN -- the companion can affect
the trajectories of the outflowing matter, concentrating it to the orbital
plane. A density gradient between equatorial and polar
regions is created, and when the star leaves AGB and the hot fast wind
starts to break its way through the
remnants of the expelled giant's envelope, the elliptical or bipolar
symmetry forms naturally.

In attempts to calculate such effects it is widely assumed, that the
intrinsic AGB wind is spherically symmetric and the asymmetry is introduced
only by the companion's influence. But this need not to be the case for
relatively close binaries, where the giant is noticeably distorted by tidal
forces. Differences in local conditions through the stellar surface, mainly
in temperature and gravity, may lead to different intensities of the
outflow. Thus the wind would show an intrinsic directivity.

We investigate this possibility, using a simple model.

\section{The model}

We assume that:
\begin{itemize}
\item{The orbit is circular and the giant corotates with orbital motion --
hence the Roche model for the gravitational potential applies.}
\item{Stellar surface is defined by the Roche equipotential surface.}
\item{The local mass loss rate per unit area, $\dot m$, is a function of
local stellar parameters such as effective temperature, $T_{\rm eff}$, and
gravity, $g$.}
\item{The luminosity is spread uniformly over the solid angle and therefore
the effective temperature depends only on radius and inclination of a given
surface element.}
\end{itemize}

We have calculated sequences of models, representing stars filling
increasing fraction of their Roche lobes for various mass ratios.
For each point of the stellar surface we have computed gravity
and effective temperature (relative to the spherical case). This allowed us
to evaluate local mass loss rates, using the prescription
derived by Arndt, Fleischer, \& Sedlmayr (1997). Integration over the
whole surface leads then to obtaining total mass loss rates.

\section{Results}

Figures 1--2 present our results.
Figs.~1a--d show the local mass loss rates (according to Arndt
et al.~prescription), represented by grayscale. For each figure black
denotes maximum and white -- minimum in the mass loss rate.
The mass loss rate at the dotted line is equal to that of a single,
undistorted (i.e. spherical) star. It marks the border between the "polar"
regions where the mass loss rate is lower, and the "equatorial strip" where
it is higher than for a single star.

\begin{figure}
\plottwo{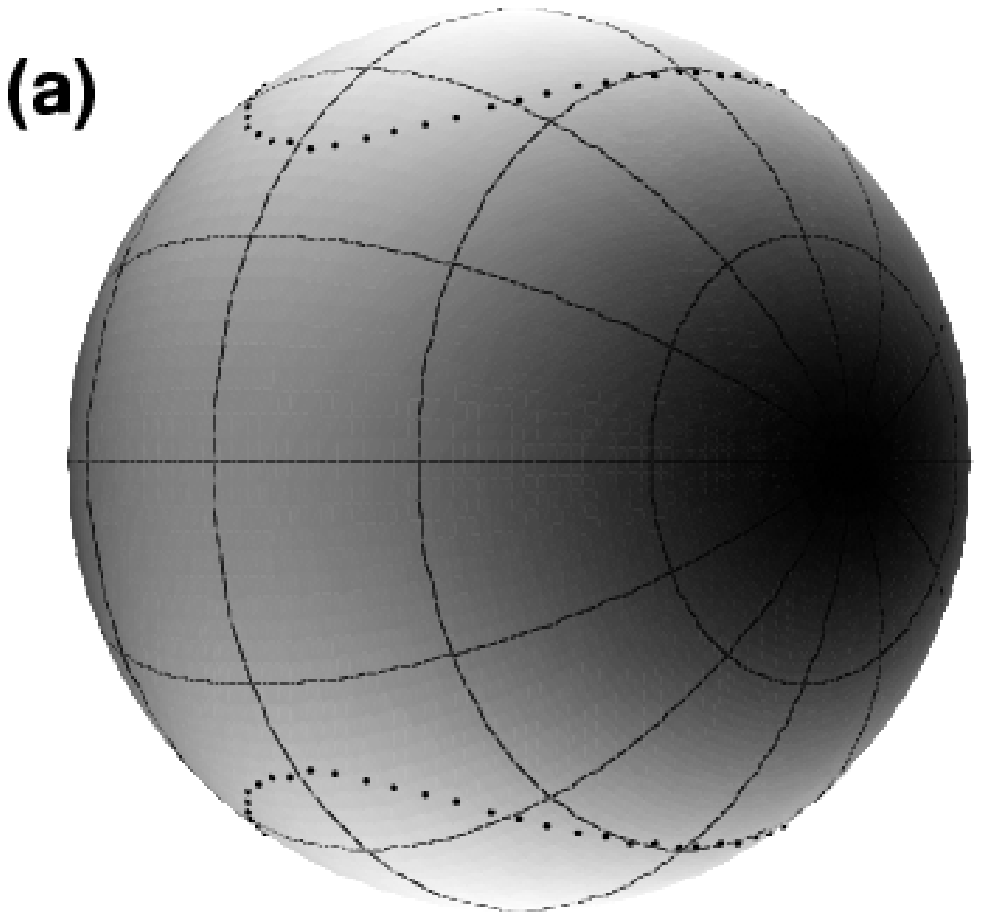}{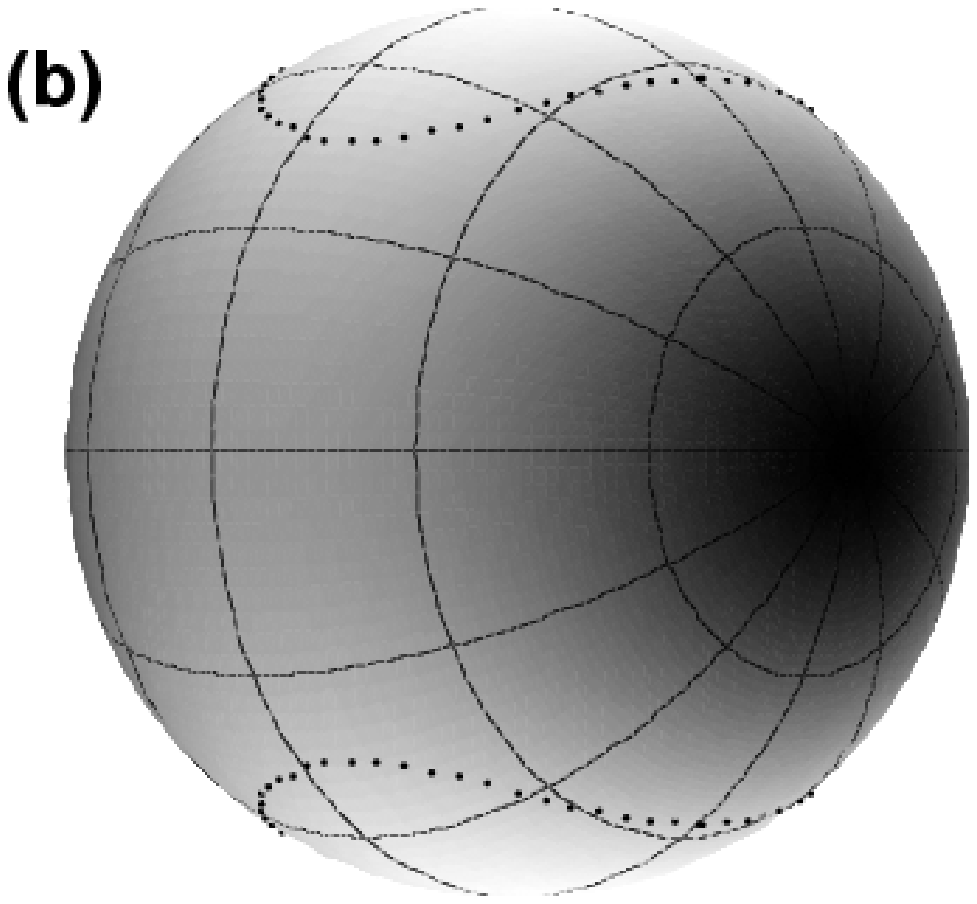}

\plottwo{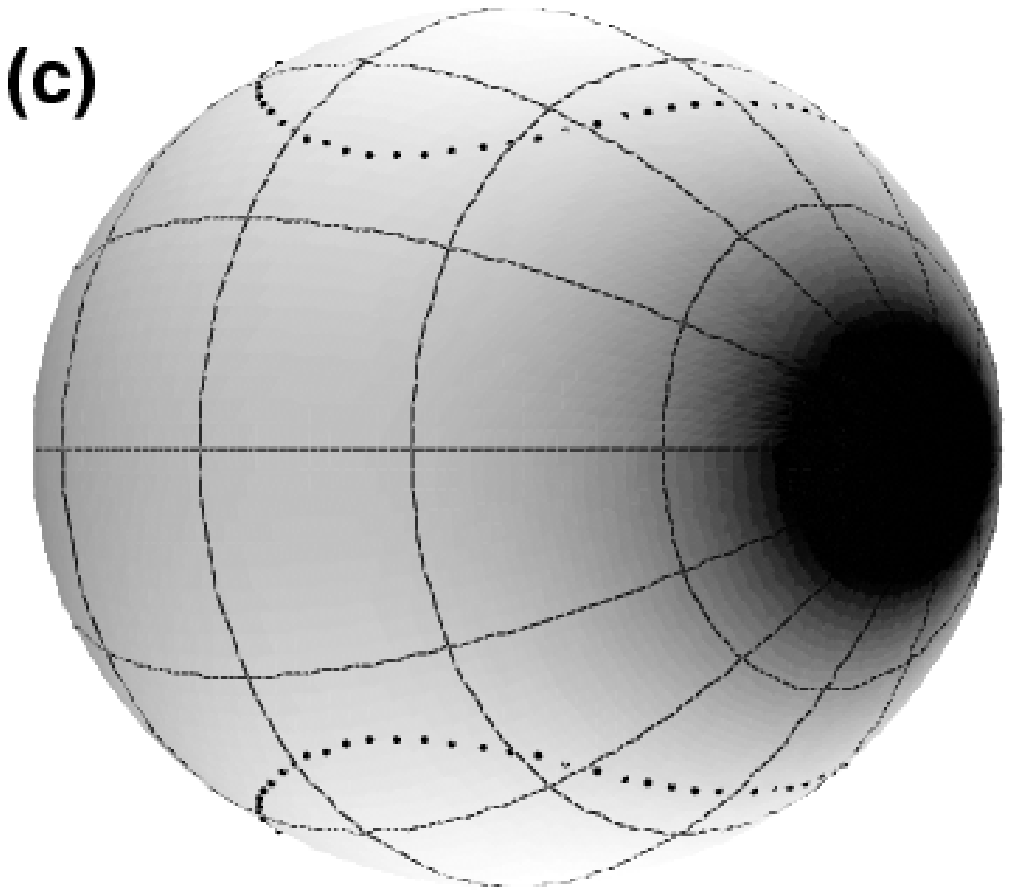}{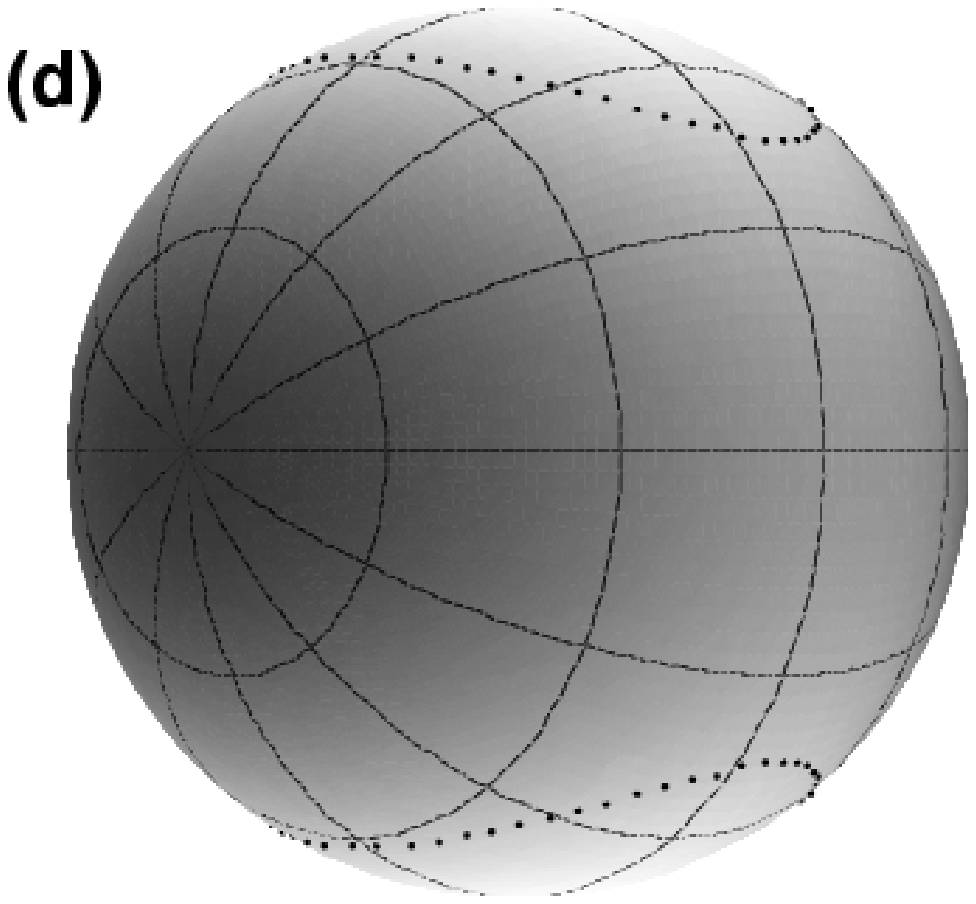}
\caption{(a) The differences in the local mass loss rate per unit area across
the surface of a giant.
Dark and light regions have, respectively, high and low mass loss rate per
unit area. At the dotted line the local mass loss rate is equal to that of
a single star.The mass ratio is $q = 0.5$.
The ratio of the radius
of the star to the Roche lobe volume radius is $r = R/R_{RL} = 0.1$.
(b) Same as Fig.~1a, but for $r = 0.5$.
(c) Same as Fig.~1a, but for $r = 1.0$.
(d) Same as Fig.~1b, but viewed from the opposite hemisphere.}
\end{figure}

In Figs.~1a--c configurations with the mass ratio $q$ = $0.5$ and the
ratio of giant volume radius to critical Roche surface volume radius
$R/R_{RL}$ = $1.0$, $0.5$, $0.1$ are shown. The stars are viewed from the
orbital plane, with the hemisphere facing the companion closer to the
observer. Fig.~1d presents for comparison the $q = 0.5$, $R/R_{RL} = 0.5$
case viewed from the opposite hemisphere.

\begin{figure}
\plotone{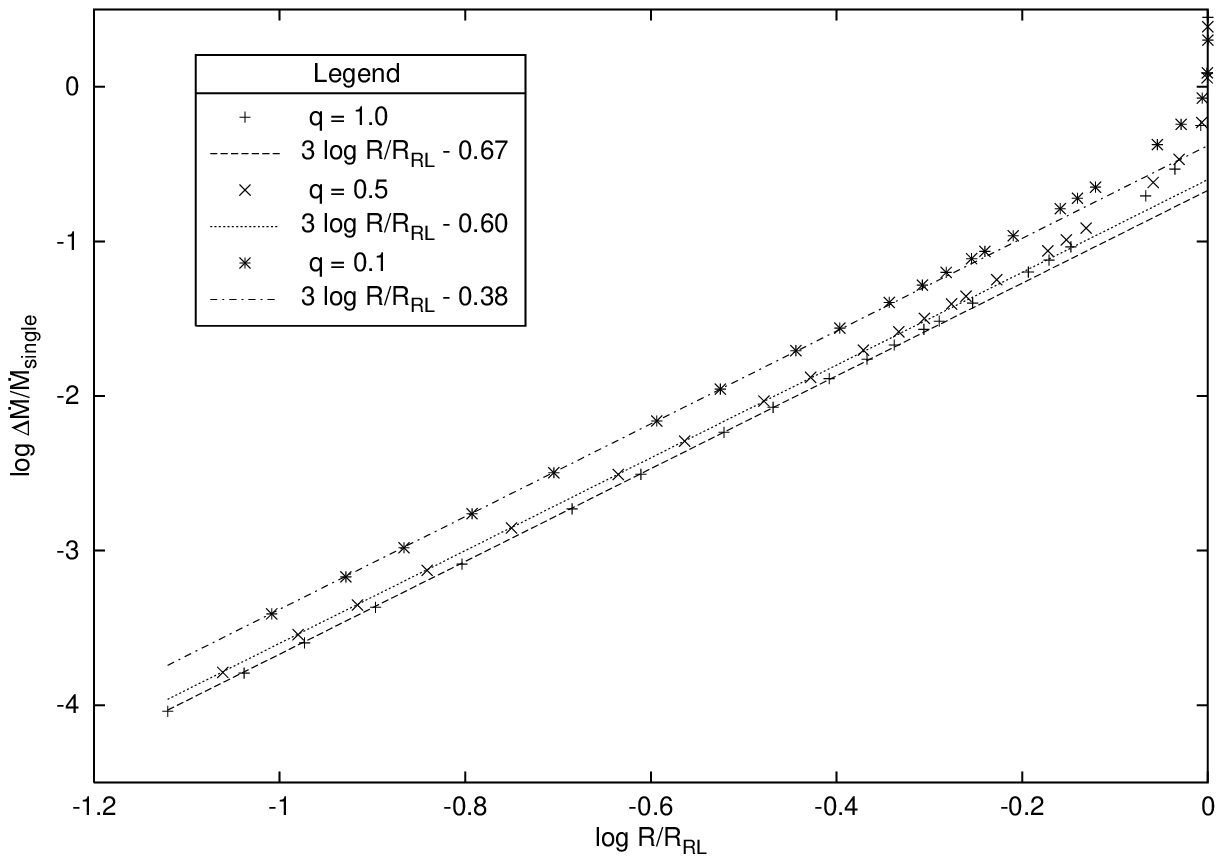}
\caption{The dependence of the total mass loss rate enhancement,
$\Delta \dot M/\dot M_{single}$, on the $R/R_{RL}$ ratio. Results for
$q = 1.0, 0.5$, and $0.1$ are shown. Points denote numerical results,
lines -- fitted linear relations.}
\end{figure}

Fig.~2 plots the enhancement of the total mass loss rate caused by the
presence of the
binary companion, $\Delta \dot M/\dot M_{single}$, against the $R/R_{RL}$
ratio.
Note logarithmic scale on both axes. Points represent numerical results,
lines -- the derived analytical relation, which is
\begin{displaymath}
\Delta \dot M/\dot M_{single} \sim (R/R_{RL})^3
\end{displaymath}
(see Sect. 4).

\section{Discussion}

As one could expect, we find significant differences in the local mass loss
rates across the surface of the giant distorted by the presence of a
companion. The gradient between the
equatorial and polar regions is evident, although the strongest enhancement
occurs toward the companion.
It would be interesting to use these results as an input for modelling
shapes of the Planetary Nebulae formed in binary systems, replacing the
assumption of intrinsic sphericity of the AGB wind.

For evolutionary calculations considering binary evolution it is important
to know by how much the stellar mass loss rate is affected by the presence
of the companion. Tout and Eggleton (1988) proposed a formula, according to
which the tidal torque would enhance the mass loss by a factor of
$1 + B \times (R/R_{RL})^{6}$, where $B$ is a free parameter to be adjusted
(ranging from $5 \times 10^{2}$ to $10^{4}$).

In our model the mass loss rate enhancement depends mainly on the $g$ value
at the giant's point closest to the binary companion.
Therefore one may try a simple analytical approach to derive
a similar relation for wind enhancement. Let us denote the gravity of
a~single star by $g_{single}$. Expanding the ratio
$g^{-1}/g_{single}^{-1}$
at the point closest to the companion
into a series in small $R/R_{RL}$ gives $const~\times~(R/R_{RL})^3$ as a
first non-zero term following the unity.
This leads to the following dependence for the mass loss rate:
\begin{displaymath}
\dot M = \dot M_{single} \, (1+ const \times (R/R_{RL})^3).
\end{displaymath}

Our numerical results confirm the above relation up to
$\log R/R_{RL} \sim -0.2$ (i.e.~$R/R_{RL} \sim 2/3$), which is shown on
Fig.~2.

\acknowledgments
This work has been supported from the grant No. 2.P03D.020.17 of the Polish
State Commitee for Scientific Research.

\end{document}